\documentclass{emulateapj}
\usepackage{apjfonts}

\newcommand{\Msun}      {\mbox{$\rm\,M_{\mathord\odot}$}}

\begin{document}

\lefthead{Neutron Star Masses with Astrometry}
\righthead{Tomsick \& Muterspaugh}

\submitted{To appear in the Astrophysical Journal}

\def\lsim{\mathrel{\lower .85ex\hbox{\rlap{$\sim$}\raise
.95ex\hbox{$<$} }}}
\def\gsim{\mathrel{\lower .80ex\hbox{\rlap{$\sim$}\raise
.90ex\hbox{$>$} }}}

\title{Masses of Neutron Stars in High-Mass X-ray Binaries with Optical Astrometry}

\author{John A.~Tomsick\altaffilmark{1} and
Matthew W.~Muterspaugh\altaffilmark{2,3}}

\altaffiltext{1}{Space Sciences Laboratory, 7 Gauss Way, University of California, 
Berkeley, CA 94720-7450 (e-mail: jtomsick@ssl.berkeley.edu)}

\altaffiltext{2}{Department of Mathematics and Physics, College of Arts and
Sciences, Tennessee State University, Boswell Science Hall, Nashville, TN
37209}

\altaffiltext{3}{Tennessee State University, Center of Excellence in
Information Systems, 3500 John A. Merritt Blvd., Box No.~9501, Nashville, TN
37209-1561}

\begin{abstract}

Determining the type of matter that is inside a neutron star (NS) has 
been a long-standing goal of astrophysics.  Despite this, most of the 
NS equations of state (EOS) that predict maximum masses in the range 
1.4--2.8\Msun~are still viable.  Most of the precise NS mass measurements
that have been made to date show values close to 1.4\Msun, but a reliable
measurement of an over-massive NS would constrain the EOS possibilities.
Here, we investigate how optical astrometry at the microarcsecond level
can be used to map out the orbits of High-Mass X-ray Binaries (HMXBs), 
leading to tight constraints on NS masses.  While previous studies by 
Unwin and co-workers and Tomsick and co-workers discuss the fact that 
the future {\em Space Interferometry Mission} should be capable of 
making such measurements, the current work describes detailed simulations 
for 6 HMXB systems, including predicted constraints on all orbital 
parameters.  We find that the direct NS masses can be measured to an 
accuracy of $\sim$2.5\% (1-$\sigma$) in the best case (X~Per), to 
$\sim$6.5\% for Vela~X-1, and to $\sim$10\% for two other HMXBs.

\end{abstract}


\section{Introduction}

Neutron stars are found in a large variety of astrophysical settings, 
such as in accreting binaries, degenerate binaries, supernova remnants, 
and as isolated objects.  They are one possible endpoint to the evolution 
of massive stars as well as being the locations of the highest magnetic
field strengths and highest densities in the Universe.  They are often
found through their radio, X-ray, or gamma-ray pulsations, which can be 
on time scales as short as milliseconds, and they can be accretion- or
rotation-powered \citep{tml93,bildsten97,abdo09}.  Despite attempts to 
determine their fundamental properties using numerous techniques, the 
form of matter that exists within a neutron star is still a mystery.

The work that has been done to try to determine the neutron star composition
has strong contributions from both theory and observation.  For an assumed
neutron star (NS) composition, the theoretical pressure-density relationship 
(i.e., equation of state, EOS) directly predicts a specific NS mass-radius
relationship \citep{lp01}, so that measurements of mass and/or radius 
provide direct constraints on the NS EOS.  It is currently even unknown 
whether NSs might be made of quark matter, which gives a radically different
mass-radius relationship from normal matter \citep{lp04}.  Most notably, 
each theoretical EOS has a maximum NS mass that can be supported \citep{lp04}, 
so that one measurement of an over-massive NS ($\sim$2\Msun~rather than the 
more canonical 1.4\Msun) would severely constrain the possibilities for the 
NS EOS.

Although accurate NS mass measurements have been made for NSs in binary 
systems, the very accurate measurements have found values close to 
1.4\Msun~\citep{tc99,lp07}.  However, more recent observations have shown
evidence for over-massive NSs with best estimates in the 1.8--2.8\Msun~range.
These have come both from High-Mass X-ray Binaries \citep[HMXBs,][]{barziv01,clark02}
and observations of binary radio pulsars \citep{freire08}.  In addition, 
there is evidence from studies of ``type I'' X-ray bursts, which are caused 
by thermonuclear flashes on the NS surface \citep{lewin93}, that the maximum 
NS mass is in the 1.9--2.3\Msun~range \citep{slb10}.  Improving the mass 
constraints and enlarging the sample of NSs with accurate mass measurements 
could finally lead to definitively ruling out EOSs.

Currently, the HMXBs with the best NS mass estimates use X-ray pulsations
to measure the projected size of the NS orbit and optical spectroscopy to
measure the radial velocities of the companion star and, thus, to constrain
the projected size of the companion's orbit \citep{cc06}.  However, even in 
cases where these measurements are possible, the binary inclination is still 
a major source of uncertainty in the measurement of the NS mass.  Thus, in
this work, we explore how NS mass measurements can be improved via optical
astrometry.  In many cases, HMXB orbits are as large as 20 to several hundred
$\mu$as, and the companion's orbit, including the binary inclination, can
be mapped directly with astrometry on the microarcsecond level 
\citep{unwin08,tsp09}.  Here, we look at the specific case of the future
{\em Space Interferometry Mission} and perform detailed simulations to
determine how well this mission would be able to constrain NS masses.

\section{Selecting Sources for Neutron Star Mass Measurements}

\begin{table*}
\caption{Estimates of System Parameters and Astrometric Signatures\label{tab:signature}}
\begin{minipage}{\linewidth}
\begin{center}
\scriptsize
\begin{tabular}{lccccccc} \hline \hline
    &   &  $d$\footnote{The source distance given in \cite{lvv06} and references therein. 
If \cite{lvv06} give a range of values, then the value we adopt in this work is given in parentheses.  
If \cite{lvv06} give multiple values, then the value we adopt comes from the literature.  Superscripts 
indicate the following references:
($aa$)\cite{steele98};
($bb$)\cite{kaper06}}
& $P_{orb}$\footnote{From \cite{lvv06} and references therein.}
& $M_{comp}$\footnote{The estimate for the mass of the companion star from the literature.  If a paper indicates
a range of values, then the value we adopt in this work is given in parentheses.  Superscripts indicate the 
following references: 
($1$)\cite{no01};
($2$)\cite{negueruela99}; 
($3$)\cite{clark01};
($4$)\cite{wg98};
($5$)\cite{kaaret99};
($6$)\cite{cox00};
($7$)\cite{barziv01};
($8$)\cite{vandermeer07};
($9$)\cite{ferrigno08};
($10$)\cite{kaper06};
($11$)\cite{on01};
($12$)\cite{rbh92};
($13$)\cite{clark00};
($14$)\cite{cm02};
($15$)\cite{ckm05};
($16$)\cite{verrecchia02};
($17$)\cite{wilson03};
($18$)\cite{negueruela03};
($19$)\cite{coe88}
($20$)\cite{reig04}}
& $M_{NS}$\footnote{We assume a neutron star mass of 1.4\Msun~unless there is a value given in the literature.
Superscripts indicate the following references:
($w$)\cite{barziv01};
($x$)\cite{quaintrell03};
($y$)\cite{kaper06};
($z$)\cite{vandermeer07}}
& $\rho$       & $\rho_{\rm comp}$\\
Source Name   & $V$     & (kpc)          & (days)     & (\Msun)           & (\Msun) & ($\mu$as) & ($\mu$as)\\ \hline
4U 0352+30/X Per           & 6.6  & 0.7--1.3 (1.0) & 250.3      & 15.5$^{3}$           & 1.4     & 1993      & 165.1\\
3A 0535+262/V725 Tau       & 9.6  & 2.0$^{aa}$     & 111         & 8--22$^{4}$ (15)     & 1.4     & 574      & 49.0\\
XTE J1946+274              & 16.9 & 9.5            & 169.2       & 10--16$^{16,17}$ (13) & 1.4     & 153      & 14.9\\
Vela X-1/GP Vel            & 6.9  & 1.9            & 8.96   & 23.8$^{7}$ & 1.86--2.27$^{w,x}$ (2.0) & 131 & 10.2\\
GX 301--2/BP Cru           & 10.8 & 3.0--4.0$^{bb}$ (3.5)       & 41.59 & 43$^{10}$        & 1.85$^{y}$ & 238 & 9.8\\
2S 1417--624               & 17.2 & 1.4--11.1 (6.0) & 42.12     & 12$^{11}$         & 1.4      & 93.7 & 9.8\\
EXO 2030+375/V2246 Cyg     & 19.7 & 7.1             & 46.02     & 17.5$^{19,6}$     & 1.4     & 94.2 & 7.0\\
SAX J0635.2+0533           & 12.8 & 2.5--5.0 (3.8) & 11.2      & 11--17$^{5,6}$ (14) & 1.4    & 64.1      & 5.8\\
EXO 0331+530/BQ Cam        & 15.4 & 7              & 34.25      & 20$^{2}$          & 1.4     & 81.8      & 5.4\\
KS 1947+300                & 14.2 & 10.0            & 40.4      & 17.5$^{18,6}$     & 1.4     & 61.3 & 4.5\\
4U 0115+634/V635 Cas       & 16.3 & 7--8 (7.5)     & 24.3       & 18$^{1}$          & 1.4     & 58.8      & 4.2\\
1E 1145.1--6141/V830 Cen   & 13.1 & 8              & 14.4      & 10$^{9}$         & 1.4     & 32.6 & 4.0\\
SAX J2103.5+4545           & 14.2 & 6.5             & 12.68     & 20.0$^{20}$       & 1.4     & 45.4 & 3.0\\
4U 1907+09                 & 16.4 & 5               & 8.38      & 26--27.9$^{15}$ (27) & 1.4   & 49.2 & 2.4\\
4U 1538--52/QV Nor         & 14.4 & 4.5--6.4 (5.5)  & 3.73      & 19.8$^{12,13}$    & 1.4      & 23.7 & 1.6\\
XTE J1855--026             & $\sim$15 & 10          & 6.07      & 25$^{14,6}$       & 1.4      & 19.4 & 1.0\\
Cen X-3/V779 Cen           & 13.3     & 9              & 2.09      & 20.2$^{8}$         & 1.34$^{z}$ & 9.9 & 0.6\\ \hline
\end{tabular}
\end{center}
\end{minipage}
\end{table*}

The best targets for obtaining neutron star mass measurements by mapping 
out binary orbits with optical astrometry should meet several criteria.  
They should be optically bright, and the angular sizes of their orbits 
should be large.  In addition to the orbit being large, the cleanest 
orbital measurements will be obtained in cases where all or nearly all 
of the optical light comes from one of the binary components.  The targets 
should be known to harbor NSs, and evidence for this can either come from 
the detection of X-ray pulsations or the presence of type I X-ray bursts.  
Finally, it is useful to know the orbital periods for planning the 
observations, and orbital periods as well as other parameters, such as
the source distance and the companion mass, must be constrained in order
to estimate the angular sizes of the orbits.

While one could consider High-Mass X-ray Binaries (HMXBs) and
Low-Mass X-ray Binaries (LMXBs), it is clear that the former
will provide the best opportunities.  A large number of HMXBs
are bright ($V\sim 5$--20), most of the optical light comes 
from the companion star, and their long orbital periods (days
to a year) mean that the separation between the NS and the
companion is large.  In addition, many HMXBs are X-ray pulsars
and for a large fraction of these the projected size of the
NS orbit ($a_{x}\sin{i}$, where $i$ is the binary inclination) 
has been measured, which, when combined with astrometric 
measurements, allows for a direct NS mass measurement.

Thus, in selecting sources, we have started with the catalog of 
Galactic HMXBs compiled by \cite{lvv06}.  This catalog includes 
114 HMXBs, including 38 sources for which both orbital periods 
and NS spin periods have been measured.  These HMXBs range from 
being very bright in the optical ($V = 6$) to not having known 
optical counterparts, and we further cut the list down to 27 
sources by requiring that the sources have a $V$-magnitude 
brighter than 20.  While all of these sources are known X-ray 
pulsars, $a_{x}\sin{i}$ has not been measured in all cases.  
As this parameter is required for a direct NS mass
measurement, we finally select the 17 HMXBs that meet the
above criteria and also have a measurement of the projected
size of the NS orbit.

The 17 sources are listed in Table~\ref{tab:signature} along 
with their $V$-band magnitudes and the information relevant 
to estimating the angular size of the companion's orbit.  If 
the semi-major axis of the binary is $a$, then the corresponding
angle subtended at the source distance $d$ is $\rho = \tan^{-1}{(a/d)}$, 
which is given by
\begin{equation}
\rho = 2.35~\mu{\rm as}~d_{\rm kpc}^{-1}~P_{\rm orb, hr}^{2/3}~(M_{\rm comp}+M_{\rm NS})^{1/3}~~~,
\end{equation}
where $d_{\rm kpc}$ is the distance to the source in kpc, 
$P_{\rm orb, hr}$ is the orbital period in hours, and
$M_{\rm comp}$ and $M_{\rm NS}$ are the masses of the companion
and the NS, respectively, in units of \Msun.  Then, the 
companion's orbital signature is given by 
$\rho_{\rm comp} = \rho M_{\rm NS}/(M_{\rm NS}+M_{\rm comp})$.  In 
Table~\ref{tab:signature}, the values of $d$ and $P_{\rm orb}$
are taken from \cite{lvv06}.  For systems where estimates of
$M_{\rm NS}$ and $M_{\rm comp}$ have been given in the literature, 
that value is given along with the reference.  Otherwise, 
companion masses are estimated using the spectral type (or 
range of spectral types) given in \cite{lvv06} and the tables 
of stellar masses given in Chapter 15 of \cite{cox00}, and 
NS masses are assumed to be 1.4\Msun.  If some of these systems 
actually contain over-massive neutron stars, then $\rho_{\rm comp}$
will be larger than our estimates in these cases.

\section{Simulations}

\subsection{The Path of the HMXB Photocenter}

To represent motions of the target photocenters on the sky, we define a 
linear coordinate system such that $x = \Delta\alpha\cos{\delta}$ and 
$y = \Delta\delta$ where $\alpha$ and $\delta$ are, respectively, the 
Right Ascension and Declination of the sources at any time, $t$.  The 
following equations represent the motion of the target photocenters
\begin{equation}
x = x_{0} + (t-t_{0})u_{x} + \pi\cos{\delta_{\mathord\odot}}\sin{(\alpha_{\mathord\odot}-\alpha_{0})} + x_{\rm orbital}~~{\rm and}
\end{equation}
\begin{equation}
y = y_{0} + (t-t_{0})u_{y} + \pi\cos{\delta_{\mathord\odot}}\sin{\alpha_{0}}\cos{(\alpha_{\mathord\odot}-\alpha_{0})} + y_{\rm orbital}~~~,
\end{equation}
where quantities with subscript ``0" are evaluated at time $t_{0}$, 
$\alpha_{\mathord\odot}$ and $\delta_{\mathord\odot}$ are the R.A. and Decl. 
for the sun, and $x_{\rm orbital}$ and $y_{\rm orbital}$ represent the 
terms in the $x$- and $y$-directions due to orbital motion.  Following
\cite{lindegren97} and section 2.3.4 of Volume 1 of the Hipparcos Catalog 
\citep{perryman97}, the orbital terms are
\begin{equation}
x_{\rm orbital} = B X(t) + G Y(t)~~{\rm and}~~y_{\rm orbital} = A X(t) + F Y(t)~~~,
\end{equation}
where $A$, $B$, $F$, and $G$ are the Thiele-Innes elements \citep{green85}, 
which are a linearized alternative to the standard Campbell orbital elements,
and depend on the angle subtended by the semi-major axis of the photocenter, 
$\rho_{\rm phot}$, the argument of periastron, $\omega$, the position angle 
of the node, $\Omega$, and the binary inclination, $i$.  

The time-dependent parameters are given by
\begin{equation}
X(t) = \cos{E} - e~~{\rm and}~~Y(t) = (1 - e^{2})^{1/2} \sin{E}~~~,
\end{equation}
where $e$ is the orbital eccentricity, and $E$ is the eccentric anomaly, 
which is given by
\begin{equation}
\frac{2\pi}{P}(t - T) = E - e\sin{E}~~~,
\end{equation}
where $P$ is the orbital parameter, and $T$ is the time of periastron 
passage.  

\begin{figure}
\includegraphics[clip,scale=0.45]{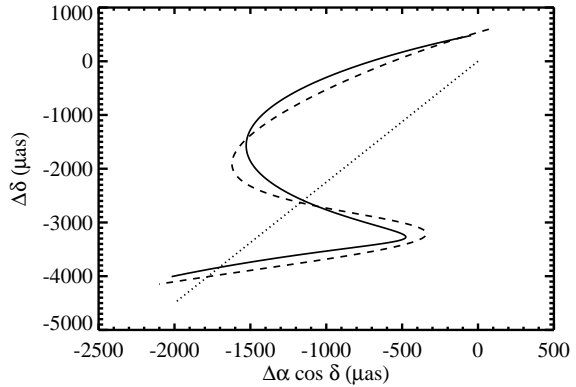}
\vspace{0.0cm}
\caption{The best estimate for the path of the companion star and
photocenter of the High-Mass X-ray Binary X~Per over a period of
1 year.  The {\em dotted line} shows the motion due only to the 
proper motion, and the {\em dashed line} shows the motion due to 
proper motion and parallax, assuming a distance of 1000~pc.  The
{\em solid line} includes these two effects as well as orbital 
motion.  For X~Per, the orbital period is 250 days, and we estimate
that the semi-major axis of its orbit will subtend an angle of
165~$\mu$as.\label{fig:path}}
\end{figure}

Thus, the path of the photocenter is determined by 12 parameters:
2 position, 2 proper motion, 1 parallax, and 7 orbital.  As mentioned
above, almost all of the light from an HMXB comes from the high-mass
stellar companion, placing the photocenter at the location of the
companion star \citep{coughlin10}.  In the following, we make the 
approximation that 100\% of the light comes from the companion so that 
$\rho_{\rm phot}$ corresponds to the semi-major axis of the companion star 
itself ($\rho_{\rm comp}$).  Figure~\ref{fig:path} shows the path of the 
companion star for the HMXB X~Per over a period of one year, illustrating 
the different contributions to the motion.

\subsection{{\em SIM Lite} Astrometry Measurements}

Here, we consider the specific capabilities of {\em SIM Lite} in 
order to carry out our simulations.  {\em SIM Lite} will measure
positions of grid stars across the sky, which will provide global
astrometric measurements to a limiting accuracy of 4~$\mu$as 
(1-$\sigma$), and it will use reference stars around selected 
targets (such as the HMXBs) to provide a co-ordinate system 
covering a smaller region ($\sim$2$^{\circ}$) where relative
astrometric measurements will have a limiting accuracy of 
1~$\mu$as (1-$\sigma$), and these are referred to as ``Narrow
Angle'' measurements \citep{nm09,unwin09}.  It is the relative 
uncertainty that is important for considering how accurately we 
will be able to measure the HMXB orbital parameters, and we use 
information from the {\em SIM Lite} project about the accuracy 
of these Narrow Angle measurements.  However, parallax and proper 
motion measurements require global astrometry. 

In addition, for our simulations, we consider the fact that the 
{\em SIM Lite} measurements provide 1-dimensional positions.  
While the path of the companion star is fully defined by $x$ and 
$y$, one must consider the baseline orientation for each measurement.
We use the angle $\theta$ to define the orientation of the baseline, 
where $\theta$ is measured from the x-axis in the counter-clockwise 
direction.  The quantity that is actually measured is angle between
the position of the target after it is projected onto the baseline
and the origin of the x-y co-ordinate system, which is given by
$r = x\cos{\theta} + y\sin{\theta}$.  For the observing campaigns
described below with many measurements, a baseline angle is chosen
at random for each observation.

Finally, for the simulations, we determine the uncertainty in each
measurement using the on-line {\em SIM Lite} ``Differential Astrometry 
Performance Estimator'' 
(DAPE)\footnote{See http://mscws4.ipac.caltech.edu/simtools\_v2/.}.
This tool allows the user to put in information about their target
as well as details about the {\em SIM Lite} observations, and it
determines a ``single measurement accuracy'' (SMA), which we use
for our simulations, as well as the amount of time required for
the observation.  An example of the input and output from this 
tool is shown in Table~\ref{tab:dape}.  Once the SMA has been 
determined for a given target, a random number generator is
used to select a number from a Gaussian distribution with $\sigma$ 
equal to the SMA, and adding this number to $r$ results in the 
measured angle $r_{meas}$.  Thus, each visit to the target generates 
the following information:  $r_{meas}$, the uncertainty in $r_{meas}$, 
which is equal to the SMA, a timestamp ($t$), and a baseline 
angle ($\theta$).

\begin{table}[b]
\caption{Differential Astrometry Performance Estimator for {\em SIM Lite}:  
Input and Output \label{tab:dape}}
\begin{minipage}{\linewidth}
\begin{center}
\begin{tabular}{lc} \hline \hline
Quantity & Value\\ \hline
\multicolumn{2}{c}{Input parameters} \\ \hline
Science target $V$-magnitude & 7\\
Science target $B-V$ color index & 0.4\\
Characteristic reference target $V$-magnitude & 10\\
Characteristic reference target $B-V$ color index & 0.4\\
Target-reference sky separation & $1^{\circ}$\\
Number of reference targets & 4\\
Number of chop cycles & 4\\
Science target integration time per chop & 30 s\\
Reference target integration time per chop & 30 s\\ \hline
\multicolumn{2}{c}{Output parameters} \\ \hline
Single measurement accuracy & 1.47 $\mu$as\\
Science target integration time per visit & 480 s\\
Reference target integration time per visit & 480 s\\
Overhead time per visit & 480 s\\
Total mission time per visit & 1,440 s\\ \hline
\end{tabular}
\end{center}
\end{minipage}
\end{table}

\subsection{Fitting for Orbital Motion}

Once the simulated data are produced, they are fitted with a function 
that accounts for proper motions and parallaxes for the target and 
reference stars as well as orbital parameters for the target.  The 
functional form is $f = x\cos{\theta} + y\sin{\theta}$, where $x$ 
and $y$ come from Equations 2 and 3 above.  In total, the equation
includes 11 free parameters:  five are non-orbital ($x_{0}$, $y_{0}$, 
$u_{x}$, $u_{y}$, $\pi$), and six are related to the binary orbit
($\rho_{\rm comp}$, $\omega$, $\Omega$, $i$, $e$, and $T$).  For the targets 
of interest, the orbital period is known very accurately, and we fix
this parameter to the known value rather than leaving it as a free
parameter.

We wrote code in Interactive Data Language (IDL) to perform the fits, 
and used the IDL routine {\em curvefit}.  This routine uses the 
Marquardt method \citep{br92}, which incorporates a gradient-expansion
algorithm and computes a least-squares fit.  It is necessary to 
carry out the computations using double-precision arithmatic with
stringent convergence criteria.  Although {\em curvefit} includes 
an option to estimate rather than exactly compute the function 
derivatives for each iteration, we obtained better results when
computing the exact derivatives.

\subsection{Cross-Check with Other Simulation Software}

We checked the software described above by comparing it to the 
code used for simulations of {\em SIM Lite} observations of 
planets \citep{traub10}.  The planet code was developed as 
part of the {\em SIM} Double Blind Test and uses realistic 
cadences and baselines but assumes 2-dimensional observations.  
There were 5 teams involved in the Double Blind Test, including
one team led by one of us (MWM).  Table~\ref{tab:comparison}
shows that there is, in general, excellent agreement between 
the two sets of code (the planet code and the new code written 
for this work).  Although there are a couple differences, they
are easy to understand.  One difference is that the proper 
motions are significantly more accurate for the planet code, 
and this is due to the fact that the simulated observations 
are spread over 5 years rather than 1 year.  Second, the other 
parameters all have slightly smaller uncertainties with the 
planet code because the SMAs were smaller in a previous version 
of the DAPE calculator. 

\begin{table}
\caption{Comparison of Measurement Uncertainties for Vela~X-1\label{tab:comparison}}
\begin{minipage}{\linewidth}
\begin{center}
\begin{tabular}{ccccc} \hline \hline
           &              &       & Planet Code & New Code\\
Parameter  &  Description & Units & Uncertainty\footnote{\scriptsize These are the 68\% confidence
(1-$\sigma$) uncertainties on the Vela X-1 parameters calculated with the planet code that was written for the {\em SIM} Double Blind Test.  These simulations include 100 observations made over 5 years.  The observations are 2-dimensional, and the single measurement accuracy in each dimension is 1.27 $\mu$as.}         & Uncertainty\footnote{\scriptsize These are the 68\% confidence uncertainties on the Vela X-1 parameters calculated with the new code written for this work.  These simulations include 200 observations made over 1 year.  The observations are 1-dimensional, and the single measurement accuracy is 1.47 $\mu$as.}\\ \hline
$e$        & Eccentricity &  ---     & $\pm$0.04 & $\pm$0.041\\
$i$        & Inclination  & degrees & $\pm$1.14 & $\pm$1.29\\
$\omega$   & Argument of Periastron & degrees & $\pm$20   & $\pm$27\\
$t_{0}$     & Time of Periastron Passage & days    & $\pm$0.5  & $\pm$0.67\\
$\Omega$   & Position angle of the node & degrees & $\pm$1    & $\pm$1.4\\
$\rho_{\rm comp}$    & Semi-major axis & $\mu$as &  $\pm$0.2  & $\pm$0.22\\
$d$        & Distance & pc     & $\pm$0.5\footnote{\scriptsize Although this is the error from the fit to the Narrow Angle observations, the actual uncertainty on the distance will depend on the tie-in to the absolute reference frame.}   & $\pm$0.7$^{c}$\\
$x_{0}$    & Reference Position & arcsec  & $\pm$$4\times 10^{-7}$ & $\pm$$5.4\times 10^{-7}$\\
$y_{0}$    & Reference Position & arcsec  & $\pm$$4\times 10^{-7}$ & $\pm$$5.4\times 10^{-7}$\\
$v_{x}$    & Proper Motion & arcsec/yr & $\pm$$1.1\times 10^{-7}$ & $\pm$$5.2\times 10^{-7}$\\
$v_{y}$    & Proper Motion & arcsec/yr & $\pm$$1.1\times 10^{-7}$ & $\pm$$5.8\times 10^{-7}$\\ \hline
\end{tabular}
\end{center}
\end{minipage}
\end{table}

\section{Results}

\subsection{HMXBs for Simulations and Observing Parameters}

\begin{figure*}[ht]
\begin{center}
\includegraphics[clip,scale=0.8]{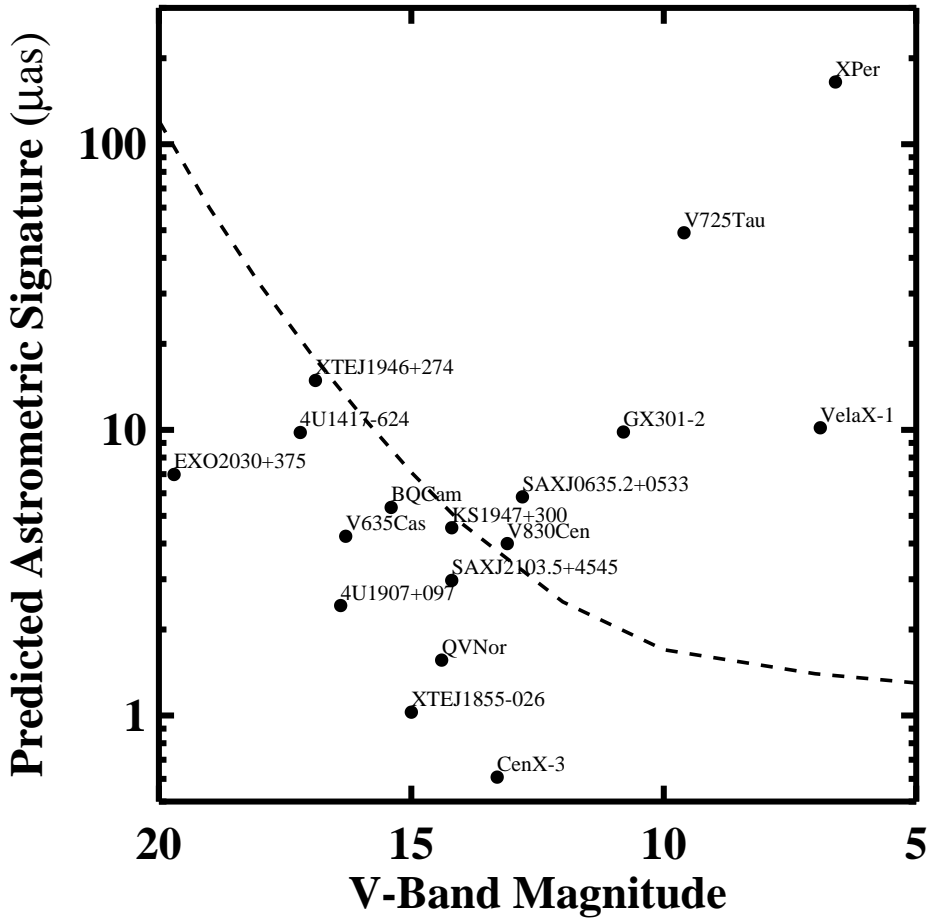}
\vspace{0.0cm}
\caption{Predicted astrometric signatures from orbital motion vs.
$V$-band magnitude for 17 neutron star HMXBs for which the projected
size of the neutron star's orbit ($a_{x}\sin{i}$) has been measured
via X-ray pulsations.  The dashed line shows the threshold for 
detection of orbital motion in 40 hours of {\em SIM Lite} mission
time.  As described in \citep{tsp09}, the threshold is defined as 
the level at which the system's semi-major axis is 10 times larger 
than the astrometry noise per observation (i.e., the single-measurement 
accuracy) divided by the square root of the number of observations.
\label{fig:signature}}
\end{center}
\end{figure*}

Figure~\ref{fig:signature} shows the predicted orbital astrometric
signatures and the $V$-band magnitudes for the 17 HMXBs listed in
Table~\ref{tab:signature}, comparing their location in the plot to 
the approximate sensitivity limit for {\em SIM Lite} taken from 
\cite{tsp09}.  The limit is based on 40~hours of mission time and 
is conservative in the sense that it corresponds to a signal-to-noise 
ratio of 10.  Thus, while some of the sources that are close to the 
limit may still have orbits that are measurable by {\em SIM Lite}, 
in this work, we focus on the six systems that are clearly above the 
line: X~Per, V725~Tau, Vela~X-1, GX~301--2, SAX~J0635.2+0533, and 
V830~Cen.

For each of our targets, we used the DAPE tool (see above) to 
determine a possible plan for observing campaigns that would use 
48~hours of mission time (per target).  We used the parameters shown 
in Table~\ref{tab:dape} except for the science target $V$-magnitude, 
which depends on the target, and the science target integration time 
per chop\footnote{In making a measurement of an angle between sources, 
the {\em SIM Lite} instrument ``chops'' back-and-forth between the 
target and the reference star by changing the length of the delay line
\citep{unwin09}.}.  For the brightest four of our six targets, we were 
able to obtain low SMAs ($<$1.8~$\mu$as) with the integration time set 
to 30~s (see Table~\ref{tab:obs}).  For the fainter two targets, we
increased the integration time to 60~s, giving SMAs of 2.39 and
2.66~$\mu$as for SAX~J0635.2+0533 and V830~Cen, respectively.
The trade-off is having smaller numbers of observations; however, 
because we still use 30~s for the reference target integration time
per chop and about a third of the time for the campaign is overhead
(see Table~\ref{tab:dape}), we only need to decrease the number of
visits in the observing campaign from 120 to 90 for the fainter two 
targets.

\begin{table}[b]
\caption{Observing Parameters for a Campaign with 48 Hours of Mission Time\label{tab:obs}}
\begin{minipage}{\linewidth}
\begin{center}
\scriptsize
\begin{tabular}{lcccc} \hline \hline
                           &           & Science Target & Single        &        \\ 
                           &           & Integration Time  & Measurement   & Number of            \\
Source Name                & $V$       & Per Chop (s)\footnote{This value was varied according to the $V$-magnitude of the source, but the other input parameters were left at the values shown in Table~\ref{tab:dape}.} & Accuracy ($\mu$as)\footnote{These values come from the Differential Astrometry Performance Estimator for {\em SIM Lite}}  & Observations\footnote{In each case, the number of observations was adjusted to obtain a total mission time, including overheads, of 48 hours.} \\ \hline
4U 0352+30/X Per           & 6.6       & 30  & 1.45 & 120\\
3A 0535+262/V725 Tau       & 9.6       & 30  & 1.56 & 120\\
Vela X-1/GP Vel            & 6.9       & 30  & 1.47 & 120\\
GX 301--2/BP Cru           & 10.8      & 30  & 1.76 & 120\\
SAX J0635.2+0533           & 12.8      & 60  & 2.39 & 90\\
1E 1145.1--6141/V830 Cen   & 13.1      & 60  & 2.66 & 90\\ \hline
\end{tabular}
\end{center}
\end{minipage}
\end{table}

While detection of orbital motion may still be possible for several 
of the other HMXBs close to the sensitivity line in 
Figure~\ref{fig:signature}, such a detection is not feasible in the 
extreme cases.  For a source as faint as EXO~2030+375 ($V = 19.7$), 
even if one increases the science target integration time per chop 
to 2 hours (the maximum allowed by the DAPE tool), the SMA is 
10~$\mu$as, which is larger than the 7~$\mu$as size of the orbit
(see Table~\ref{tab:signature}).  Although one could conceivably
still detect the orbit with a large number of visits, even as few 
as 20 visits would require nearly 1 month of mission time.  Thus,
detecting the orbit would likely take several months and would 
clearly not be a good use of {\em SIM Lite} time.

\subsection{Fitting Results}

For each of the six HMXBs, we simulated 2,000 {\em SIM Lite} observing 
campaigns and fitted the data with the 11 parameter model described 
above (note that there are 11 rather than 12 free parameters because
we always fixed the orbital period to the known value).  For each 
HMXB and each parameter, we produced histograms of the 2,000 best fit 
values obtained.  For all of the parameters of all of the HMXBs except 
for V830~Cen, the histograms appear to be normally distributed, allowing 
us to fit each histogram with a Gaussian and use the $\sigma$ value to 
estimate the uncertainties on each parameter that {\em SIM Lite} will 
be able to obtain.  For V830~Cen, the histogram for the eccentricity 
parameter is not normally distributed.  We suspect that the parameter 
is not well-constrained because V830~Cen has the smallest estimated 
semi-major axis, $\rho_{\rm comp} = 4.0$~$\mu$as, and we discuss the case of
V830~Cen further below.

We report the simulation results for the other five HMXBs in 
Table~\ref{tab:results1}, including the six orbital parameters 
as well as the parallax measurement.  {\em SIM Lite} will greatly
improve constraints on the binary inclinations, $i$.  While the
current uncertainties are on the order of $\sim$10$^{\circ}$, with
a large systematic component, {\em SIM Lite} will constrain $i$
to within $\pm$0.28$^{\circ}$, $\pm$0.98$^{\circ}$, $\pm$1.58$^{\circ}$, 
and $\pm$2.50$^{\circ}$ for X~Per, V725~Tau, Vela~X-1, and
GX~301--2, respectively.  Also, {\em SIM Lite} will provide the 
first direct measurements of the angular size of the companion's 
semi-major axis, $\rho_{\rm comp}$.  For example, for X~Per and Vela~X-1, 
our simulations indicate measurements of $165\pm 0.45$~$\mu$as and 
$10.2\pm 0.28$~$\mu$as.

For each HMXB, Table~\ref{tab:results1} provides a measurement of the 
parallax, $\pi$, and then an actual measurement of the distance.  The 
uncertainty shown with $\pi$ is that obtained from the fit to the data.
However, there is also an uncertainty from the fact that the reference
frame will only be known to within 4~$\mu$as, and both error components
are given with the distance measurement.  If X~Per is at a distance of
1000~pc, we predict that {\em SIM Lite} will measure its distance to
an accuracy of $\sim$5~pc.

\begin{table*}
\caption{Simulation Results and Mass Measurements (all parameters free)\label{tab:results1}}
\begin{minipage}{\linewidth}
\begin{center}
\scriptsize
\begin{tabular}{lccccc} \hline \hline
Source Name & X Per & V725 Tau & Vela X-1 & GX 301--2 & SAX J0635.2+0533\\ \hline
\multicolumn{6}{c}{{\em SIM Lite} orbital measurements}\\ \hline
$e$             & $0.11\pm 0.0026$ & $0.47\pm 0.0099$ & $0.0898\pm 0.050$ & $0.462\pm 0.068$ & $0.29\pm 0.15$\\
$i$ ($^{\circ}$) & $29.82\pm 0.28$ & $30.50\pm 0.98$ & $82.52\pm 1.58$ & $67.40\pm 2.50$ & $49.12\pm 7.84$\\
$\omega$ ($^{\circ}$) & $108\pm 1.5$ & $310\pm 2.2$ & $332.59\pm 33.6$ & $130.4\pm 8.2$ & $356\pm 36$\\
$t_{0}$ (days) & $\pm$0.99 & $\pm$0.33 & $\pm$0.84 & $\pm$0.84 & $\pm$0.88\\
$\Omega$ ($^{\circ}$) & $45\pm 0.5$ & $45\pm 2.0$ & $45\pm 1.6$ & $45\pm 2.7$ & $45\pm 12$\\
$\rho_{\rm comp}$ ($\mu$as) & $165\pm 0.45$ & $49.0\pm 0.41$ & $10.2\pm 0.28$ & $9.8\pm 0.49$ & $5.8\pm 0.56$\\ \hline
\multicolumn{6}{c}{{\em SIM Lite} distance measurement}\\ \hline
$\pi$ ($\mu$as) & $1000\pm 0.87$ & $500\pm 0.37$ & $526\pm 0.28$ & $278\pm 0.31$ & $263\pm 0.68$\\
$d$ (pc) & 1000($\pm$0.87)($\pm$4) & 2000($\pm$1.5)($\pm$16) & 1900($\pm$1.0)($\pm$14) & 3500($\pm$4.0)($\pm$52) & 3800($\pm$9.9)($\pm$58)\\ \hline
\multicolumn{6}{c}{X-ray measurement of the projected size of the pulsar's orbit}\\ \hline
$a_{x}\sin{i}$ (lt-s)\footnote{These are from the following:
($1$)\cite{delgado01};
($2$)\cite{bildsten97};
($3$)\cite{kaaret00}} & $454.0\pm 4.0$$^{1}$ & $267\pm 13$$^{2}$ & $113.89\pm 0.13$$^{2}$ & $368.3\pm 3.7$$^{2}$ & $83\pm 11$$^{3}$\\ \hline
\multicolumn{6}{c}{Derived neutron star mass measurement and error contributions}\\ \hline
$M_{NS}$ (\Msun) & $1.40\pm 0.035$ & $1.40\pm 0.15$ & $2.00\pm 0.13$ & $1.85\pm 0.21$ & $1.40\pm 0.59$\\ 
$\delta$$M_{NS,distance}$ (\Msun)   & 0.008 & 0.014 & 0.018 & 0.032 & 0.029\\
$\delta$$M_{NS,\rho_{\rm comp}}$ (\Msun)      & 0.004 & 0.014 & 0.062 & 0.103 & 0.156\\
$\delta$$M_{NS,i}$ (\Msun)         & 0.025 & 0.087 & 0.110 & 0.176 & 0.463\\
$\delta$$M_{NS,a_{x}\sin{i}}$ (\Msun) & 0.023 & 0.125 & 0.004 & 0.036 & 0.329\\ \hline
\end{tabular}
\end{center}
\end{minipage}
\end{table*}

\begin{table*}
\caption{Simulation Results and Mass Measurements ($e$ and $\omega$ fixed)\label{tab:results2}}
\begin{minipage}{\linewidth}
\begin{center}
\scriptsize
\begin{tabular}{lcccc} \hline \hline
Source Name & Vela X-1 & GX 301--2 & SAX J0635.2+0533 & V830 Cen\\ \hline
\multicolumn{5}{c}{{\em SIM Lite} orbital measurements}\\ \hline
$e$             & 0.0898\footnote{Fixed.} & 0.462$^{a}$ & 0.29$^{a}$ & 0.20$^{a}$ \\
$i$ ($^{\circ}$) & $82.52\pm 1.55$ & $67.40\pm 2.29$ & $49.12\pm 7.78$ & $60.68\pm 9.87$\\
$\omega$ ($^{\circ}$) & 332.59$^{a}$ & 130.4$^{a}$ & 356$^{a}$ & 128$^{a}$\\
$t_{0}$ (days) & $\pm$0.040 & $\pm$0.27 & $\pm$0.31 & $\pm$0.47\\
$\Omega$ ($^{\circ}$) & $45\pm 1.6$ & $45\pm 2.4$ & $45\pm 10$ & $45\pm 13$\\
$\rho_{\rm comp}$ ($\mu$as) & $10.2\pm 0.27$ & $9.8\pm 0.37$ & $5.8\pm 0.52$ & $4.0\pm 0.56$\\ \hline
\multicolumn{5}{c}{{\em SIM Lite} distance measurement}\\ \hline
$\pi$ ($\mu$as) & $526\pm 0.27$ & $278\pm 0.31$ & $263\pm 0.67$ & $125\pm 0.51$\\ 
$d$ (pc) & 1900($\pm$1.0)($\pm$14) & 3500($\pm$4.0)($\pm$52) & 3800($\pm$9.7)($\pm$58) & 8000($\pm$33)($\pm$256)\\ \hline
\multicolumn{5}{c}{Derived neutron star mass measurement and error contributions\footnote{The value of $a_{x}\sin{i}$ used for V830~Cen is $99.4\pm 1.8$ lt-s \citep{rc02}.  The values used for the other sources are the same as in Table~\ref{tab:results1}.}}\\ \hline
$M_{NS}$ (\Msun) & $2.00\pm 0.13$ & $1.85\pm 0.19$ & $1.40\pm 0.59$ & $1.40\pm 0.57$\\
$\delta$$M_{NS,distance}$ (\Msun)     & 0.018 & 0.032 & 0.029 & 0.062\\
$\delta$$M_{NS,\rho_{\rm comp}}$ (\Msun)        & 0.061 & 0.078 & 0.145 & 0.242\\
$\delta$$M_{NS,i}$ (\Msun)           & 0.108 & 0.161 & 0.463 & 0.516\\
$\delta$$M_{NS,a_{x}\sin{i}}$ (\Msun)  & 0.004 & 0.036 & 0.329 & 0.044\\ \hline
\end{tabular}
\end{center}
\end{minipage}
\end{table*}

For Vela~X-1 and GX~301--2, the X-ray pulsation measurements 
\citep{bildsten97} give values of $e$ and $\omega$ that are much more 
accurate than our simulations predict for {\em SIM Lite}.  Thus, we
have re-fit our simulated data for these two systems and also for 
the two fainter sources, SAX~J0635.2+0533 and V830~Cen, keeping 
$e$ and $\omega$ as fixed parameters.  Once this is done, the V830~Cen
histograms appear to be normally distributed.  Table~\ref{tab:results2}
shows the fitting results for these four systems.  The improvements in 
the constraints on the most important parameters ($i$, $\rho_{\rm comp}$, and
$d$) are relatively small.  The largest improvement is in the measurement
of the time of periastron passage ($t_{0}$), which is expected since
this parameter is related to $e$.

\subsection{Direct Neutron Star Mass Measurements}

It is possible to write an expression for the mass of a neutron star, 
$M_{NS}$, in an HMXB in terms of five directly measurable quantities.
Two of the quantities: $P_{\rm orb}$ and the projected linear size of 
the neutron star orbit ($a_{x}\sin{i}$); are measured by instruments
other than {\em SIM Lite}.  The other three quantities: $\rho_{\rm comp}$, 
$i$, and $d$; will be measured by {\em SIM Lite} as described above.  
Starting from the standard orbital equations that are typically used 
for measurements of the components of binary systems 
\citep[e.g., Equation 5.3 in][]{cc06}, we derive the following equation
\begin{equation}
M_{NS} = \frac{4\pi^{2}}{G P_{\rm orb}} \frac{d~\tan{\rho_{\rm comp}}}{\sin^{2}{i}} 
[(a_{x}\sin{i}) + d~\tan{\rho_{\rm comp}}~\sin{i}]^{2}~~~,
\end{equation}
where $M_{NS}$, $P_{\rm orb}$, $d$, and $a_{x}\sin{i}$ are in the same units
as the gravitational constant, $G$, e.g., CGS units, and $i$ and $\rho_{\rm comp}$ 
are angles.  For X-ray pulsars, the projected size of the neutron star 
orbit can often be determined to very high accuracy, e.g., $a_{x}\sin{i} = 
113.89\pm 0.13$ light-seconds for Vela~X-1 \citep{bildsten97}. By obtaining 
long (years) time base-line X-ray or optical observations, X-ray pulsar 
orbital periods are, for our purposes, known with negligible uncertainties.  

In addition to the {\em SIM Lite}-measured parameters in the top parts of
Table~\ref{tab:results1}, the values of $a_{x}\sin{i}$ are given for the
5 HMXB systems, and this, in turn, is used to calculate $M_{NS}$.  For
X~Per, our simulations predict that {\em SIM Lite} will measure the 
neutron star mass to $\sim$2.5\% ($M_{NS} = 1.40\pm 0.035$~\Msun, where
the uncertainty corresponds to the 68\% (1-$\sigma$) confidence level).  
The second-best measurement would be obtained for Vela~X-1, for which
we predict a $\sim$6.5\% uncertainty on $M_{NS}$.  It should be noted
that these values consider both components (from the fit and from the
reference frame) of the distance uncertainty.  For each HMXB, the 
contribution from each parameter ($d$, $\rho_{\rm comp}$, $i$, and $a_{x}\sin{i}$) 
to the uncertainty on $M_{NS}$ is given in Tables~\ref{tab:results1} and 
\ref{tab:results2}.  For V725~Tau, the largest contribution to the uncertainty 
comes from the $a_{x}\sin{i}$ term, so that the future measurement of $M_{NS}$ 
can be substantially improved with X-ray observations of this system.

\section{Discussion and Conclusions}

The results of these simulations show that {\em SIM Lite} will provide
excellent constraints on the masses of neutron stars in HMXBs.  There is
currently no direct method for obtaining the binary inclinations ($i$) 
of HMXBs, and this will be a major improvement in the measurements.  This
is especially true for non-eclipsing systems like X~Per and V725~Tau, 
for which $i$ is thought to be near $30^{\circ}$, but it is currently 
only estimated to about $\pm$10--15$^{\circ}$.  With such a large uncertainty
in $i$, we do not have any current estimate of the NS masses in these
systems, which is why we assume 1.4\Msun~for the simulations.  Since
these are both accreting systems, there is a strong possibility that 
the NS masses are significantly higher, which could allow for a constraint
on the NS EOS.

The current constraints on $i$ are better for Vela~X-1, which is an 
eclipsing system, but HMXBs with relatively short orbital periods
can be eclipsing for a wide range of inclinations; thus, the improvement
in the measurement of $i$ that {\em SIM Lite} will provide is important.
There are already suggestions that the Vela~X-1 NS is over-massive, and
if {\em SIM Lite} finds that the NS mass is, e.g., $2.00\pm 0.13$~\Msun, 
as found with our simulations, this would rule out many NS EOSs \citep{lp04}.

There are also possibilities for further improvements to the accuracy 
of the NS mass measurements estimated in Tables~\ref{tab:results1} and
\ref{tab:results2}.  We assume in our simulations that we will use 
{\em SIM Lite} to determine the times of periastron ($t_{0}$), but 
if contemporaneous X-ray observations can be made, it will be possible
to accurately determine this time using the X-ray pulsations.  Also, 
we find that, after {\em SIM Lite} measurements of V725~Tau, the 
uncertainty in $M_{NS}$ will be dominated by the measurement of
$a_{x}\sin{i}$; thus, X-ray measurements to improve the measurement of
this parameter could ultimately lead to a mass constraint that is 
nearly as good as that for X~Per.  Finally, one more piece of information
that we have not considered in this work is the radial velocity 
semi-amplitude of the companion star ($K_{\rm comp}$).  This parameter is
related to the projected size of the orbit, and a cross-check between
this parameter and the astrometric parameters ($\rho_{\rm comp}$, $d$, $i$, 
and $e$) will be possible \citep[e.g.,][]{pj00}.  

Although we focus on obtaining NS masses via orbital measurements in 
this work, it should also be pointed out that {\em SIM Lite} can also
contribute to constraining EOSs by improving measurements of NS radii.
Two techniques for measuring the radii of NSs in Low-Mass X-ray Binaries
use measurements of the X-ray emission from the neutron star surface:  
one is to observe the thermal X-rays from the surface of a NS when
the accretion rate is very low \citep[e.g.,][]{rutledge02}; and another 
is to measure the thermal emission near the end of a type I X-ray
burst \citep{gop08,obg10}.  In both cases, the distance to the 
LMXB is the largest uncertainty in the NS radius measurement
\citep[see the references above and][]{tsp09}, and {\em SIM Lite} 
will provide accurate distances for a large number of LMXBs.

Finally, it is important to note that our requirement in this work
that $a_{x}\sin{i}$ be a measured parameter eliminates many interesting
HMXBs that will be excellent targets for {\em SIM Lite}.  In 
\cite{tsp09}, we find $\sim$20 HMXBs for which the predicted astrometric
signatures will be large enough for orbital measurements.  This includes
well-known sources such as the black hole system Cyg~X-1 \citep{cn09}
and the likely black hole system SS~433 \citep{blundell08} as well as 
systems where it is not clear whether the compact object is a black hole
or a NS.  These include the interesting case of 4U~1700--377, which is
thought to be a NS based on its X-ray properties, but which has a
compact object mass estimated at 2.4\Msun.  Also, precise compact object 
masses have not been determined for the gamma-ray binaries 
LS~I+61$^{\circ}$~303 and LS~5039 \citep{dubus06}, but given that there 
are only a small number of high-mass binaries known to produce gamma-ray 
emission, determining the type of compact object in these systems is of 
great interest.

\acknowledgments

This work was sponsored in part by the National Aeronautics and 
Space Administration (NASA) through a contract with the Jet Propulsion 
Laboratory, California Institute of Technology.  MWM acknowledges 
support from the Townes Fellowship Program and the State of Tennessee 
Centers of Excellence program.  The {\em SIM} planet-finding code was 
developed as part of the {\em SIM} Double Blind Test with support from
NASA contract NAS7-03001 (JPL\#1336910).  JAT acknowledges useful 
communications with Valeri Makarov about the planned operation of 
{\em SIM Lite} in narrow-angle mode.  JAT acknowledges useful discussions
with Sabine Reffert, Andreas Quirrenbach, Stuart Shaklan, Xiaopei Pan, 
and Shri Kulkarni.


\end{document}